\documentclass{iaus}
\usepackage{graphicx}
\title[Bright end of the CMR]{Development of the Red Sequence \\in Galaxy Clusters}%
\author[Noelia Jim\'enez et al.]{Noelia Jim\'enez$^{1}$,
Sof\'ia A. Cora$^{1}$, \\Anal\'ia Smith Castelli$^{1}$ \and 
Lilia P. Bassino$^{1}$}

\affiliation{$^1$ Facultad de Ciencias Astron\'omicas y Geof\'isicas
de La Universidad Nacional de La Plata and Instituto de Astrof\'isica de La Plata (CCT La Plata, CONICET, UNLP)\\ email: {\tt njimenez@fcaglp.unlp.edu.ar}}

\pubyear{2009}
\volume{267}  
\pagerange{}
\date{?? and in revised form ??}
\jname{Co-Evolution of Central Black Holes and Galaxies}
\editors{B.M.\ Peterson, R.S.\ Somerville, \& T.\ Storchi-Bergmann, eds.}
\begin{document}

\maketitle

\begin{abstract}
  We investigate the origin of the colour-magnitude relation (CMR)
observed in cluster galaxies by using a combination of a cosmological
{\em N}-body simulation of a cluster of galaxies and a semi-analytic
model of galaxy formation.  The departure of galaxies in the bright
end of the CMR with respect to the trend denoted by less luminous
galaxies could be explained by the influence of minor mergers.


\end{abstract}

With the aim of understanding the development of the CMR depicted by
early-type galaxies in galaxy clusters, we apply a semi-analtytic
model of galaxy formation (\cite[Lagos et al. 2008]{Lagos08}) to a
simulated galaxy cluster with virial mass $\approx 1.3 \times 10^{15}
\, h^{-1} \, {\rm M}_{\odot}$ (\cite[Dolag et al. 2005]{Dolag05}).  The
semi-analtytic model considers gas cooling, star formation, galaxy
mergers, disc instabilities, metal enrichment and feedback from
supernovae and active galactic nuclei.  Photometric properties of
simulated early-type galaxies are compared with those of early-type
galaxies observed in the central region of the Antlia cluster
(\cite[Smith Castelli et al. 2008]{SmithCastelli08}); all magnitudes
are obtained in the Washington photometric system ($C$, $T_1$).  We
find that the general trend of simulated and observed CMR are quite
similar.  However, the more massive simulated galaxies ($-22 <M_{\rm
T_{\rm 1}}<-19$) depart from the fit to observed data, displaying an
almost constant colour ($C-T_{\rm 1}\approx 1.7)$, as detected in
other clusters.  We select galaxies in six magnitude bins within the
range $-22<M_{\rm T_{\rm 1}}<-16,$ and analyse the contribution to
the stellar component of quiescent star formation and starbursts during
disk instabilities and mergers events. Major and minor mergers are
distinguished between dry and wet ones according to the cold gas mass
of the remnant.  For the most luminous galaxies, we find that the
increase of stellar mass at low redshift arises as a consequence of
minor dry merger events, in agreement with \cite[Skelton et
al. (2009)]{Skelton09}.  Galaxies lying in the faint end of the CMR
($-17<M_{\rm T_{\rm 1}}<-16$) mainly increase their stellar mass as a
result of disk instabilities. This seems to indicate that the effect
of minor dry mergers would yield to an increase of the luminosity
without strongly affecting the galaxy colours.  We will deepen this
investigation focusing on the chemical enrichment history of the
accreted stellar mass with the aim of disentangling the role played by
dry mergers in determining the properties of the more massive
galaxies.

\end{document}